\documentclass[prl,superscriptaddress,twocolumn,preprintnumbers]{revtex4}
\usepackage{color}
\usepackage[active]{srcltx}
\usepackage{amsmath,amsfonts,amssymb,amsthm,amstext,amscd,eucal,srcltx}
\usepackage{epsfig,graphicx,bm}
\usepackage{epstopdf, epsf}
\usepackage{dcolumn}
\usepackage{hyperref}

\newcommand{\be}{\begin{equation}}
\newcommand{\ee}{\end{equation}}

\newcommand{\bse}{\begin{subequations}}
\newcommand{\ese}{\end{subequations}}
\newcommand{\bea}{\begin{eqnarray}}
\newcommand{\eea}{\end{eqnarray}}
\newcommand{\ba}{\begin{array}}
\newcommand{\ea}{\end{array}}
\newcommand{\bc}{\begin{center}}
\newcommand{\ec}{\end{center}}

\begin{document}
\preprint{IPM/P-2012/009} 
\vspace*{3mm}
\title {Invisible QCD as Dark Energy}%

\author{Stephon Alexander}
\email{Stephon\_Alexander@brown.edu}
\affiliation{Department of Physics, Brown University, Providence, Rhode Island 02912, USA}

\author{Antonino Marcian\`o}
\email{marciano@fudan.edu.cn}
\affiliation{Center for Field Theory and Particle Physics \& Department of Physics, Fudan University, 200433 Shanghai, China}

\author{Zhi Yang}
\email{14110190013@fudan.edu.cn}
\affiliation{Center for Field Theory and Particle Physics \& Department of Physics, Fudan University, 200433 Shanghai, China}

\begin{abstract}

\noindent We account for the late time acceleration of the Universe by extending the QCD color to a $SU(3)$ invisible sector (IQCD). If the Invisible Chiral symmetry is broken in the early universe, a condensate of dark pions (dpions) and dark gluons (dgluons) forms.  The condensate naturally forms due to strong dynamics similar to the Nambu--Jona-Lasinio mechanism.  As the Universe evolves from early times to present times the interaction energy between the dgluon and dpion condensate dominates with a negative pressure equation of state and causes late time acceleration. We conclude with a stability analysis of the coupled perturbations of the dark pions and dark gluons.

 \end{abstract}
%\date{\today}
\maketitle

\textbf{\emph{Introduction}}

\noindent
A confluence of cosmological data tell us that the Universe is currently accelerating (see {\it e.g.} \cite{Riess:1998cb} and \cite{Frieman:2008sn, Astier:2012ba} and references therein). While the data can be explained with a cosmological constant, it is also possible that the Universe is dominated by a form of Dark Energy (DE) with a negative pressure barotropic index $w \!\sim\! -1$.  Another possibility is that general relativity is modified in the infrared (IR) admitting self accelerating solutions, and such models are still under active research \cite{carroll}.  In this work we take the position that the Einstein-Hilbert action is the correct IR description of gravity and the cosmological constant is zero, by some yet unknown mechanism. We present a model where DE emerges from an Invisible QCD (IQCD) sector due to the interaction of invisible pions and gluons which were present in the early Universe.

There are two ways to theoretically motivate an Invisible QCD sector: it has been known for a long time that the $E_8\times E_8$ Heterotic Superstring Theory, both in the weak and strong coupling limit necessarily gives rise to non-abelian gauge theories that do not directly couple to the standard model $SU(3)_{c}\times SU(2)_{W}\times U(1)_{Y}$ gauge group. While String Theory has this property, it is also possible to obtain a dark copy in another way, which corresponds to modifying general relativity by enlarging the spin connection $\omega(e)^{IJ}_{\mu}$ to a larger group $G\sim SU(N)$, which breaks to $SU(2)\times SU(N-2)$ where the $SU(N-2)$ factor is identified with the dark sector. This philosophy was pursued in \cite{unif}. Here we take a phenomenological perspective and simply assume that this dark sector exists and focus on the cosmological consequences.\\

In this model, late time acceleration arises from extending the color sector of QCD to have an ``invisible-copy''.  IQCD has similar quantum field theoretic properties of QCD, in that it is confining in the IR.  It is well known that pions arise as Goldstone modes associated to Chiral Symmetry Breaking (CSB), and in turn the microphysical description of CSB, the Nambu--Jona-Lasinio mechanism \cite{Nambu:1961}, is a strongly coupled version of superconductivity induced by hard gluon exchange.  A key feature of our DE model uses the same physics of CSB in the IQCD sector. During the matter and radiation era, the dark pions and gluons have negligible effects.  However, we will show that at late times the interaction energy between the dark pions and gluons become more significant because they remain nearly constant, mimicking an effective cosmological constant.  Through the consistency of the coupled field equations, this interaction energy naturally leads to late time acceleration and we find an interesting connection between the scale of CSB and the scale of DE.       
 
\noindent

{\it{\textbf{The Theory}}} 

\noindent We assume that Chiral symmetry in the IQCD sector is broken at some scale $f_{_D}$ corresponding to the mass of a dark pion (dpion) $\pi_{_D}$ \footnote{From now on, we will remove the subscript $D$ when referring to dark pions.}. At the renormalizable level, such a hidden sector can communicate with the standard model only through gauge interactions: $SU(3)_c$, $SU(2)_{W}$ or $U(1)_Y$ \cite{CFS,KOS,hill}. In this work, for clarity of presentation, we assume no gauge coupling to the standard model.  However, this model can easily be extended such that the gauge-confined quarks are coupled vectorially to $SU(2)_W$.  The authors of Ref.~\cite{hill} showed that a dark sector with purely vectorial coupling to the standard model has remarkable universal features.  A specific parity symmetry (known as G-parity) acting only on the hidden sector fields is left unbroken and stable weakly interacting dark-pions become a compelling candidate for a dark-matter particle.  It would be interesting to revisit the constraints on the coupling to the visible sector that is simultaneously consistent with dark-matter and DE.  We find it intriguing that our model has the possibility of connecting late time acceleration to Dark Matter and will pursue this possibility in \cite{AAM}.\\ 

The $SU(N)_D$ gauge theory we are considering is assumed to have a behavior similar to QCD. The gauge coupling becomes strong in the IR limit, triggering confinement and chiral symmetry breaking at a scale $\Lambda_D$. Below $\Lambda_D$, the effective theory is described by ``pions'' which are pseudo-Nambu--Goldstone bosons (pNGBs) associated with the spontaneously broken global flavor symmetry of the hidden sector.

For concreteness and without loss of generality, we consider the subgroup, $SU(2)_{L}\times SU(2)_{R} \rightarrow SU(2)_{V}$ with its gauge field $A^a_{~\mu}$, where
$a,b=1,2,3$ and $\mu,\nu=0,1,2,3$ are for
the dark color and space-time indices, respectively. The gauge field strength $F$ is
\be\label{F-general}
F^a_{~\mu\nu}=\partial_\mu A^a_{~\nu}-\partial_\nu
A^a_{~\mu}-g\epsilon^a_{~bc}A^b_{~\mu}A^c_{~\nu}, 
\ee
where $\epsilon_{abc}$ is the totally antisymmetric Levi-Civita symbol, the structure constant of the $SU(2)$ algebra.
We are led to consider the most general gauge invariant action coupled to ``dark quarks'':
\be 
\cal{L}\rm_{IQCD}\equiv\cal{L}\rm_{A} +\cal{L}\rm_{D} =-\frac{1}{4 g^2}F^a_{~\mu\nu}F_a^{~\mu\nu} + \imath\,\overline{\psi}\, D_{\mu}\gamma^{\mu}\psi \,. 
\ee
Here $D_\mu$ stands for the covariant derivative with respect to the gravitational connection, $\gamma^\mu\!=\!\gamma^I\, e_I^\mu$ and the metric field is decomposed in tetrad, namely $g_{\mu\nu}=e^I_{\mu} \, e^I_{\nu}$, the inverse of which is denoted as $e^\mu_I$ and the internal $SO(3,1)$ indices of which are $I=1,2...4$. Given that we are working in a system where the dpion forms as a result of CSB, the decay constant $f_{_D}$ is defined through the coupling of the axial current to the dpion. In particular, dpions can be created by the axial isospin currents. 

Matrix elements of $\mathcal{J}_{\, a}^{5 \, I}(x)$ between the vacuum and an on-shell dpion state can be parametrized as (see {\it e.g.} \cite{Peskin:1995ev})
\bea 
\label{current} 
\langle0|\mathcal{J}_{ \,a }^{5\, I }(x)|\pi^b(p)\rangle = - \imath\, \delta_a^b\, p^{I}\, f_{_D}\, e^{-\imath p \cdot x}\,, 
\eea
where the chiral current is $\mathcal{J}_{\, a}^{5\, I}(x) \!=\! \overline{\psi}(x)\,\gamma^I\gamma_{5}\,\tau_a\,\psi(x)$ and $|\pi(p)\rangle$ is the pion condensate state with normalization $\langle \pi(p')|\pi(p)\rangle = 2(2\pi)^{3}p^{0}\delta^{3}(p'-p)$.  Relation (\ref{current}) can be recast in terms of the dpion field as
\bea \label{expJ}
\langle  0| \mathcal{J}_{ \,a }^{5\, I }(x) | \pi^b \rangle=f_{_D}\, \delta^b_a \,\partial^I \pi_b(x) \,.
\eea
We can rotate the expectation value of  $\langle  0| \mathcal{J}_{ \,a }^{5\, I }(x) | \pi^b \rangle$ within the internal Lorentz indices' space, so to accomplish an explicitly space-like axial vector with vanishing temporal component. The symmetry of the vacuum state on the FLRW background allows to further reduce \eqref{expJ} to a homogenous axial vector:
\bea \label{expJ2}
\langle  0| \mathcal{J}_{ \,a }^{5\, I }(x) | \pi^b \rangle=f^2_{_D}\, 
\delta_{a}^i \,\pi(t, 0) \,,
\eea
where $\pi(t)\!\equiv\! || \pi^a(t)||$, with respect to the internal indices. The interaction of the axial current with the gauge field ${\cal L}^{ int.}_\pi = g \,A^{a}_{\mu}\,\mathcal{J}_{a}^{5\, \mu}$ is therefore consistent with homogenous and isotropic metrics.

The low energy dpion effective Lagrangian reads
\bea \label{pio}
\cal{L}\rm_{\pi}^0 = -\frac{1}{2} \partial_{\mu}\pi_a\partial^{\mu}\pi^a +\frac{\lambda}{4}\left(\pi^{a}\,\pi_{a}  - f^2_{D}\right)^{2}\,.
\eea
Consequently, the total effective Lagrangian reads
\begin{eqnarray} \label{lato}
{\cal L}_{ tot.} \!\!\!\!&=& \!\! {\cal L}_{\rm GR} +{\cal L}_{\rm A}+{\cal L}^{ int.}_\pi +{\cal L}^0_{\pi} \\
&=& \!\!\! M^2_p \,R \rm-\frac{1}{4} F^a_{~\mu\nu}F_a^{~\mu\nu} \! + g \,A^{a}_{\mu}\,\mathcal{J}_{a}^{5\, \mu} + \cal{L}\rm_{\pi}^0, \nonumber
\end{eqnarray}
in which we have introduced the reduced Planck mass as $M_p^2= (8 \pi G)^{-1}$. Quark fields have been integrated out in the path integral in order to get the effective action. The interaction term ${\cal L}^{ int.}_\pi =  g \,A^{a}_{\mu}\,\mathcal{J}_{a}^{5\, \mu}$, which entails parity violations of the $SU$(2) subgroup of the dark sector, preserves renormalizability. The total action is ${\cal S}_{tot.}= \int d^4x \,e\,\, {\cal L}_{tot.}$, with $e\!=\!\sqrt{-g}$ denoting the determinant of the tetrad $e_\mu^I$.

\noindent 

{\it{\textbf{Field Equations}}}

\noindent Solutions to the field equations that are consistent with a FLRW background can be recovered once a rotationally invariant configuration for the gauge field has been implemented: 
\be\label{A-ansatz-background}
A^a_{~\mu}=\left\{
\begin{array}{ll} a(t)\,\phi(t)\,\delta^a_i\, ,\qquad  &\ \ \mu=i\,,
\\
0\,, \qquad &\ \ \mu=0\,.
\end{array}\right.
\ee
Thanks to \eqref{expJ} and \eqref{A-ansatz-background}, the energy-momentum tensor of the theory is isotropic and homogenous. In particular, the energy-momentum tensor associated to the interaction Lagrangian yields the remarkable feature of having a barotropic index $w=-1$, since energy and pressure densities respectively read 
\begin{eqnarray}
\rho_{_{\bf AJ}}\!\!&=& 3\,  g\, f_D^2 \, \phi(t) \,\pi(t), \nonumber \\
-\, P_{_{\bf AJ}} \!\!&=&3 \,g \, f_D^2\, \phi(t) \, \pi(t), \nonumber
\end{eqnarray} 
$g$ denoting above the absolute vale of the coupling constant.
This naturally leads towards a de Sitter accelerating phase of the Universe, as soon as the interaction term becomes dominant.

In the rotationally symmetric configuration \eqref{A-ansatz-background} the gauge field-strength's components simplify and read $F_{0i}^a = \partial_{t}(\phi(t)a(t)\delta^{a}_{i})$ and $F_{ij}^{a}=-g \,\epsilon_{ij}^{a}(\phi(t)a(t))^{2}$, having specified our system in co-moving coordinates $ds^2\!=\!dt^2\!-\!a^2(t)\,d\vec x^2$. Using (\ref{A-ansatz-background}), the total gauge Lagrangian becomes 
\be
\!\!\mathcal{L}^{\bf A}+\, \mathcal{L}^{\rm int.}\!= \frac{1}{2\, g^2}\left( 3 \,(\dot{\phi}+H\phi)^2 -3 \, g^2 \phi^4\right)+ 3 g \phi\, \bar{J}(a), 
\ee
where $\bar{J}(a)\!\equiv\! f^2_{_D} \pi(t)$. From $ \frac{1}{a^{3}} \frac{\partial}{\partial t}(a^{3}\frac{\partial \cal{L}}{\partial \dot{\phi}}) \!=\!\frac{\partial {\cal{L}}}{\partial \phi}$, we obtain the equation of motion for $\phi$, which captures the dynamics of $A^a_{~\mu}$ through (\ref{A-ansatz-background}), namely
\be \label{eom1}
\ddot{\phi} + 3 H \dot{\phi}+ (2 H^2 +\dot{H}) \, \phi+ 2 \, g^{2} \phi^{3}  - g \, \bar{J}(a) =0 \,.
\ee

The equation of motion for the dpion field is recovered varying (\ref{pio}), within the assumption of spatial homogeneity. This is plausible, since a previous inflationary epoch of the universe can smooth out the dpion field. In the next section, we show that the dpion field remains homogenous against perturbations. 

Using the decomposition in a homogeneous absolute value (in the internal space) times a space-dependent unit vector, {\it i.e.} $\pi_a=||\pi_a|| \,n_a=\pi(t)\, n_a(x)$, we recover for the pion field
\be \label{eom2}
\ddot{\pi} + 3H\dot{\pi} + \lambda\, \pi(\pi^{2} -f^2_{D}) - 3 \,g f_D^2 \phi  
=0\,.
\ee
To gain some insight as to why we might expect to see late time acceleration, consider the slow roll regime of the dpion field, which is obtained by neglecting the acceleration term.  In this approximation, when the dpion field exhibits an inverse scaling with time, $\pi = \pi_0\, a^{-1}(t)$, the equation of motion reduces to
\be
3H\, \frac{\dot{a}}{a^2} = \frac{ \lambda }{a} \left( \frac{\pi_0^2}{a^2}- f^2_D \right)\,.
\ee
Solving this latter results in a power law acceleration of the Universe, namely $H(t)\simeq t^{-1}$, provided that $\pi(t_0)\!=\!\pi_0\!>\!\!>\!f_{_D}$ and, as customary when taking into account cosmological scalar fields, the slow roll condition holds: $3H\dot{\pi}\simeq V'\!>\!\!>\! \ddot{\pi}$. When the interaction term $\cal{L}\rm_{\rm int} \!=\! -g\,\phi(t)\, f^2_{_D} \pi(t)$ between the dpion and the gauge field is considered, we will see that this term persists to have a nearly constant energy density yielding a negative pressure equation of state. Finally, it is straightforward to show that a slightly different behavior in the time dependence of the dpion, {\it i.e.} $\pi \!=\! \pi_0 \,a^{-n}(t)$ with $n>0$, would yield the same late time-behavior $H(t)\simeq t^{-1}$.\\  

Late time acceleration is recovered when the gauge field  asymptotically evolves in time as the scale factor, and the pion field approaches the constant value $\pi\simeq f_D$. Below we show how these self consistent solutions to the equations of motion for the pion and gauge field can be recovered, by working in comoving coordinates and assuming the expansion of the universe to be given by de Sitter phase. Finally, given the non-linearities in the coupled differential equations, we pursue a numerical analysis of the full system of equations.

First we consider the energy density $\rho$ and the pressure $P$ for our low energy effective system that emerges from the dark sector, namely
\be\label{rho-P-total} 
\rho= \rho_{_{\bf YM}}+\rho_{_{\bf AJ}}\ ,\qquad P=\frac13\rho_{_{\bf YM}}-\rho_{_{\bf AJ}}, 
\ee
where 
\be\label{rho0-rho1}
\rho_{_{\bf YM}} =\frac{3}{2}(\dot{\phi}+H \phi)^2 +\frac{3}{2}g^2\, \phi^4 \
, \qquad \rho_{_{\bf AJ}}= 3\, g\,\phi\,\bar{J}(a)\,,
\ee
and recall that the Friedmann equations are given by
\be\begin{split} \label{friedmann}
H^2 \,M^2_p &=\frac{1}{2} ( \dot{\phi}+H \phi )^2 + \frac{1}{2}g^2 \phi^4 + g\,\phi\bar{J}(a) \\
&\ \ \ + \frac{1}{6} \dot{\pi}^2 + \frac{\lambda}{12} \left( \pi^2-f_D^2 \right)^2  ,\cr
(\dot{H}+H^2)  \,M^2_p~&=-\frac{1}{2} (\dot{\phi} +H \phi)^2 -\frac{1}{2}g^2\, \phi^4+  g\, \phi \, \bar{J}(a) \\
& \ \ \  -\frac{1}{3} \dot{\pi}^2 +\frac{\lambda}{12} (\pi^2 -f_D^2 )^2 \,.
\end{split}
\ee
Having derived the differential equations that govern the dynamical system, we can now proceed to solve it. \\

{\it{\textbf{Field Dynamics.}}} 

\noindent Unlike usual gauge field theories, where the gauge fields dilute during cosmic expansion, the coupling of the gauge field to the dquark current leads to a growth of its homogenous component. This can be understood from the inspection of the equations of motion on the FLRW background. The growth of the gauge field will generically occur as it scales with $a(t)$, while the interaction energy $\rho_{_{AJ}}\!=\!g \,\phi\, \bar{J}(a)$ remains nearly constant at late times. We can then get ahead with our purpose of solving the full dynamical system for the fields involved, so as to plot the evolution of barotropic index 
\begin{eqnarray}
&&\!\!\!\!\!\!w=\frac{P}{\rho} =\frac{P_{_{\bf AJ}} + P_{_{\bf YM}}+  P_\pi}{\rho_{_{\bf AJ}}+\rho_{_{\bf YM}}+\rho_\pi  } = \\
&&\!\!\!\!\!\!= \frac{ \frac{1}{2} (\dot{\phi} +H \phi )^2 +\frac{1}{2}g^2 \phi^4- 3g\phi\bar{J}(a)  \!+\!\frac{1}{2} \dot{\pi}^2 \!-\!\frac{\lambda}{4} (\pi^2\!-\!f_D^2)^2 }{\frac{3}{2} (\dot{\phi} +H \phi )^2 + \frac{3}{2}g^2 \phi^4+ 3 g\phi \bar{J}(a)\!+\!\frac{1}{2} \dot{\pi}^2 \!+\!\frac{\lambda}{4} (\pi^2\!-\!f_D^2)^2 }\,. \nonumber
\end{eqnarray}
Under customary assumption, we are able to solve for the coupled system of differential equations in the configuration space $\{\phi(t), \pi(t) \}$, and to find solutions consistent with a de Sitter expanding phase. We assume in \eqref{eom2} the slow roll condition for $\pi$. Furthermore, we assume that at any time energy densities are dominated by the coupling of the gauge field to the axial current, $\rho_\pi\!<\!\!<\!\rho_{_{\bf AJ}}$ and $\rho_{_{\bf YM}}\!<\!\!<\!\rho_{_{\bf AJ}}$, and show later that these assumptions are consistent with the solutions obtained. In this heuristic analysis, the dynamical system is analytically solved imposing initial conditions at recombination. From \eqref{eom2} and \eqref{friedmann} we find
\be \label{solpi}
\pi(t)\simeq  f_D \left[1- c_0 \exp \left(-\frac{2 \lambda f_D^2 \,t}{3H} \right) \right]^{-\frac{1}{2}}\,,
\ee
which asymptotically reaches the value $\pi_\infty=f_D$, and in which $c_0$ is an initial constant. We can then solve for the gauge field, and within a similar assumption on its time derivatives than the condition imposed for $\pi(t)$ we find
\be \label{solphi}
\phi(t)\simeq  \frac{f_D}{(2g)^{\frac{1}{3}}}\, \left[1- c_0 \exp \left(-\frac{2 \lambda f_D^2 \,t}{3H} \right) \right]^{-\frac{1}{6}}\,.
\ee 
Both solutions \eqref{solpi} and \eqref{solphi} are monotonically decreasing and converge asymptotically towards values that are proportional to $f_D$; thus their product conspire to provide an accelerating solution well approximated by a de Sitter phase, the effective cosmological constant of which assumes the asymptotic value
\be
\Lambda \simeq \frac{f_D^4}{M_p^2} \,.
\ee
Supernovae data, which entail at current times $H\simeq 10^{-42}$~GeV, are consistent with the asymptotic value for the gauge field $A \!\simeq\! f_D\!\simeq\!~\!10^{-3}\!$ eV, the coupling constant $g$ having been assumed to be order unity. This suggests a fascinating conclusion: cosmic acceleration is the result of CSB in the dark sector, since it occurs at the same scale of energy, {\it i.e.} $M_{_{DE}}\simeq f_D$.

Conservation of the energy-momentum tensor is easily checked, and the attractor behavior of the solutions is also recovered. Indeed, specifying the numerical values $g\!=\!\lambda\!=\!0.1$, the system of differential equations provides the unique fixed point: 
\begin{equation}
(\phi_f,\pi_f, H_f)\!=\!(2.17\cdot 10^{-3} {\rm eV}, \, 2.04\cdot 10^{-3} {\rm eV},1.08 \cdot10^{-35} eV). \nonumber
\end{equation}
Upon linearization, the first order dynamical system recast in term of derivatives in the variable $N=\ln a(t)$, provides a matrix the eigenvalues of which have negative real components. This ensures that the fixed point is an attractor, and that the system asymptotically converges towards a de Sitter phase.  

Finally, a more cogent numerical analysis corroborates the analytical investigations reported above. Related plots are shown in Fig.1 and in Fig.2, which make evident the transition from a radiation dominated epoch (at the recombination $z\!\simeq\!1100$) to our present time ($z=0$) dominated by DE. 

\begin{figure}[htb] \label{figuna}
\begin{center}
\includegraphics[scale=0.72]{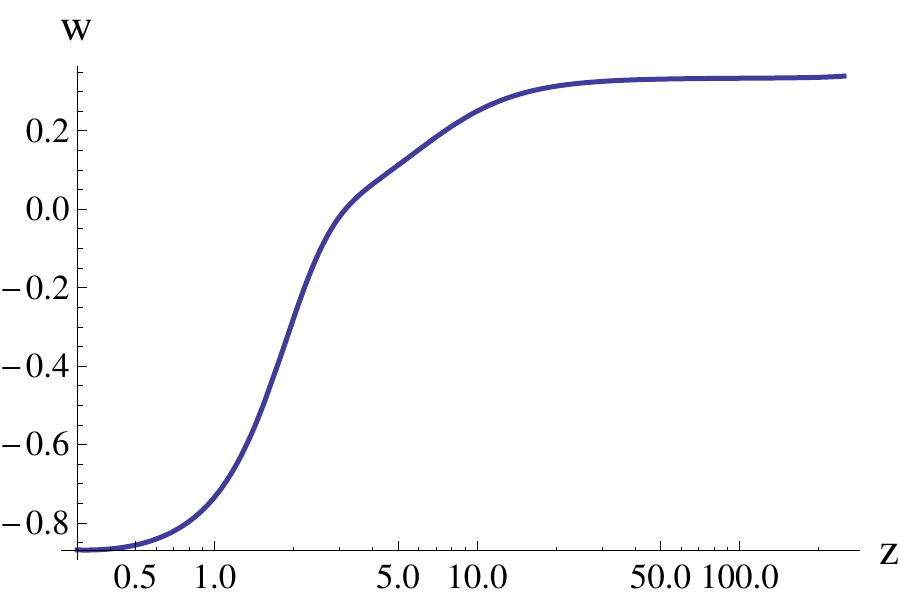}  
\caption{Plot of $w$ against the redshift $z$, from current time up to recombination. The coupled system of non-linear differential equations \eqref{eom1}, \eqref{eom2} and \eqref{friedmann} has been solved numerically for $g\!= \!10^{-1}$ and $\lambda\!=\!1$, and the initial conditions on the functions $H(0)\!=\! 10^{-42} GeV\,$ and  $\phi(0)\!\simeq\!\pi(0)\!\simeq\!f_D \!=\!10^{-3} \, eV\,$, and on their derivatives $\phi'(0)\!\simeq\! \pi'(0)\!=\! 10^{-5}\, (eV)^2\,$. Transition from DE to (dark sector) radiation happens for $z\simeq2$.}
\includegraphics[scale=0.72]{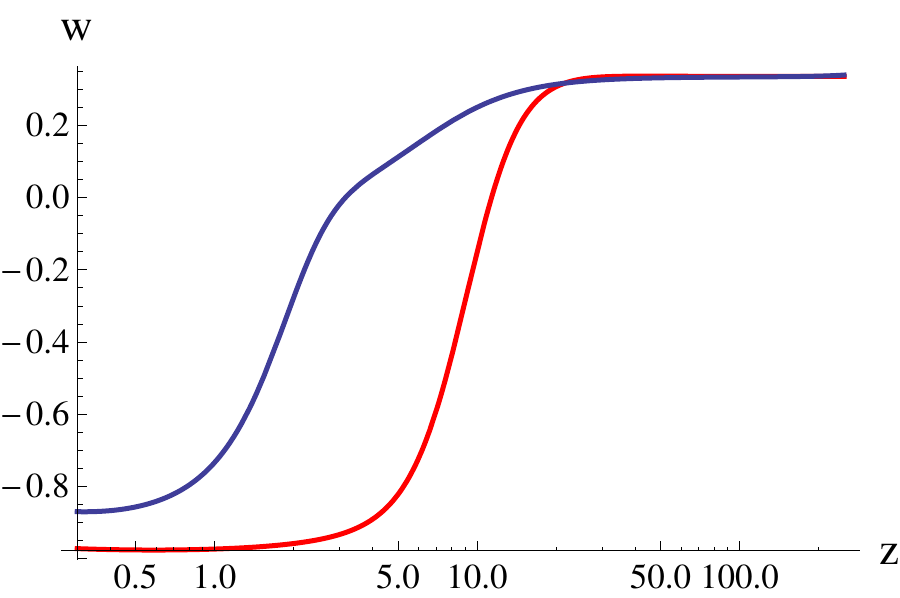}  

\caption{Plot of $w$ against $z$, as sourced only by the dark sector. The initial conditions for the coupled system of differential equations are the same as in the previous figure, but the coupling constant $g$ has been chosen an order of magnitude smaller (red line). }
\end{center}
\vspace{-4mm}
\end{figure}

\noindent 

{\it{\textbf{Perturbation Analysis.}}}

\noindent The analysis of perturbations can be developed on the FLRW background by implementing the choice of conformal coordinates $g_{\mu\nu}=a(\eta)^2\,\eta_{\mu \nu}$. Varying action (\ref{lato}), we recover the gauge field equation of motion:
\bea \label{vara}
&&\eta^{\mu \rho} (\partial_\mu F^a_{\rho \nu} + g\, \epsilon^a_{\ bc}\, A^b_{\mu} F_{\rho\nu}^c) = g\, a^4  \mathcal{J}^{5\, a}_\nu \,. \nonumber
\eea 
From the above equation, the gauge field spatial perturbations immediately follow:
\bea
\!\!\!&&\Box \delta A^a_i + g\,\left( \epsilon^a_{\ bc} A^b_j \, \partial^j \delta A_i^c + \, \epsilon^a_{\ bc}\, \delta A^b_j \, \partial^j \! A_i^c 
\right)+\nonumber \\
\!\!\!&&+ g^2\! \left( \delta A^k_b A^b_{[k} A^a_{i]} + A^k_b \delta A^b_{[k} A^a_{i]}+ A^k_b A^b_{[k} \delta A^a_{i]} \right) = \nonumber\\
\!\!\!&&= g\, f_D\, a(\eta)^4\,\, \partial_i \delta \pi^a(\vec{x}, \eta)\,.  
\eea
The time component perturbations (denoted by $'=  d/ d\eta$) of the gauge field are found to be:
\be \nonumber
\Box\, \delta A^a_0 + g \epsilon^a_{\ bc}  A^b_k \, \partial^k \delta A_0^c- g^2 A^i_b \delta A_{[0}^b\, A^a_{i]} 
= g \,a^4\, \delta{\pi^a}'(\vec{x}, \eta)\,.
\ee
As stated above while using co-moving coordinates, the background solutions for $A^a_j(\eta)$ are subjected to the gauge $A^a_j(\eta)=A(\eta)\,\delta^a_j$, which reduces the spatial perturbation equation to
\bea
\!\!\!\!\!\!&&\Box \,\delta A^a_i(\vec{x}, \eta) + g\, A(\eta)\,[ \, \nabla \wedge \delta \vec{A}(\vec{x},\eta) ]^a_i +\\
\!\!\!\!\!\!&&+\, g^2 \Big[ 2 A^2(\eta)\,  {\rm Tr}[\delta A(\vec{x},\eta)] \,\delta^a_i \, \Big] =g\, a(\eta)^4\,\, \partial_i \delta\pi^a(\vec{x}, \eta)\,,  \nonumber\\ \label{varone}
&&\Box\, \delta A^a_0 (\vec{x}, \eta)+ g\,A(\eta) \epsilon^a_{\,\,bc}  \partial^b \delta A_0^c(\vec{x}, \eta)+ \nonumber\\
&&+2 g^2 A(\eta)^2 \delta A_0^a (\vec{x}, \eta)= g\, a(\eta)^4\,\, \delta{\pi^a}'(\vec{x}, \eta)\,.
\eea
Taking the trace of the gauge-field perturbations and writing $\delta A^a_i= {\rm Tr}[\delta A^a_i(\vec{x}, \eta)] \,\delta^a_i /3$ yield
\bea \label{varto}
&\Box \,{\rm Tr}[\delta A^a_i(\vec{x}, \eta)] -6 g^2 A^2(\eta) \,{\rm Tr}[\delta A^a_i(\vec{x}, \eta)]  \nonumber\\
& = g\, a(\eta)^4\,\, {\rm Tr}[\partial_i \delta\pi^a(\vec{x}, \eta)] \,.
\eea
Implementing the same gauge, we recover the equation for the perturbations to the dpion field,
\bea \label{varpi}
\!\!\!\!\!\!\!&\frac{\Box}{a} \delta \pi^a (\vec{x}, \eta) \!+\! 2 \frac{a'}{a^2} \delta {\pi^a}'(\vec{x}, \eta) \!+\! \lambda a \,\delta {\pi}^a \!(\vec{x}, \eta)\! \left( 3 \,\pi(\eta)^2 \!-\!f^2_D \right)\! \nonumber\\
\!\!\!\!\!\!\!&= - g f_{_D}\, \left(\partial^i \delta A_i^a (\vec{x}, \eta) + \delta {A_0^a}' (\vec{x}, \eta) \right) 
\,.
\eea
We are interested in seeing if sub-horizon modes of the dpion field, (\ref{varpi}), develop instabilities and cluster. Therefore we focus on wavelengths which are either sub-horizon or above the binding energy of the dpions involved, $H\!\!>\!\!>\!|\vec{k}|\!>\!\!>\!\!f_{_D}$, by studying the evolution of plane-waves $\delta \pi(\vec{x},\eta)\!\simeq\! \alpha (\eta) \exp \,( \imath \vec{k}\!\cdot\! \vec{x})$ solutions to (\ref{varpi}).  The equation of motion further reduces to $\alpha'' + 2H a\, \alpha' +  (\vec{k}^2 + \lambda \,a^2 f^2_D) \simeq 0$, solutions of which are superpositions of spherical Bessel functions of the first and second type, 
\bea
\alpha(\eta)\!&=& \!\alpha_1\,\,  j_\nu\left(\frac{k}{Ha(\eta)}\right)    + \alpha_2 \,\,y_\nu\left(\frac{k}{Ha(\eta)}\right)\,, \nonumber \\
\nu \!&=&\! \frac{-H \pm \sqrt{H^2 - 4 \lambda f_D^2 }}{2H}\,, \nonumber
\eea
in which $k\!=\!|\vec{k}|$, and are convergent to zero at late times for a proper choice of the initial conditions, $\alpha_1\in \mathbb{R}$ and $\alpha_2\!=\!0$. Indeed, assuming $\lambda\!<\!\!<\!1$ we see that sub-horizon modes are oscillatory and bounded over all the time axis. This behavior mimics the behavior of super-horizon modes of scalar fields during inflation and is a consequence of the acceleration of space-time. Zero modes are constant and not evolving in time.  
 
Finally, perturbations of the gauge field decrease exponentially (in comoving time), {\it i.e.} show the conformal time behavior 
\bea 
{\rm Tr} [\widetilde{\delta A}^a_i(k,\eta)]\simeq  1/a(\eta)\,.  \nonumber
\eea
We therefore conclude that all sub-horizon perturbations are suppressed. In future work we will perform a full perturbation analysis taking into account metric perturbations \cite{YMR}.  

\noindent 

{\it{\textbf{Conclusion and Discussion.}}} 

\noindent We have provided a model of late time acceleration from minimal assumptions, in that aside from instantiating a dark non-abeilan copy, we have not introduced any new physics. In fact, we have employed the well known physics of the Nambu--Jona-Lasinio mechanism of chiral symmetry breaking in ordinary QCD applied to IQCD, which is well motivated from string theory.  Late time acceleration emerges from the interaction between gravity, a chiral condensate and an invisible gluon that fills the universe today.  A preliminary perturbation analysis shows that DE does not cluster on sub-horizon modes.

Finally, we leave to detailed investigations (see {\it e.g.} \cite{AAM}) the analysis of constraints on the coupling to the visible sector, and the eventual behavior as dark-matter of dquark and dpions in this scenario. Indeed we find intriguing that our model has the possibility of connecting late time acceleration to Dark Matter (see {\it e.g.} \cite{Dona:2015xia,Addazi:2016sot}).

{\it{\textbf{Acknowledgement.}}} We are indebted to David Spergel for enlightening discussions during the early stage of this work. We also thank Andrea Addazi, Robert Caldwell, John Collins, Gia Dvali, Daniel Grinstein and Diego Guadagnoli for their useful feedback. S.A. is supported by the US Department of Energy
under grant DE-SC0010386. A.M. wishes to acknowledge support by the Shanghai Municipality, through the grant No. KBH1512299, and by Fudan University, through the grant No. JJH1512105.

\end{document}